\documentclass[aps,prd,
twocolumn,
floatfix,nofootinbib,groupedaddress,showpacs,showkeys]{revtex4-1}
\usepackage{graphicx}
\usepackage{amsmath}
\usepackage{latexsym}
\usepackage{here}
\usepackage{array}
\usepackage{dcolumn}
\usepackage{bm}
\usepackage{epsfig}

\newcommand{\Ref}[1]{(\ref{#1})}
\newcommand{\beao}{\begin{eqnarray*}}
\newcommand{\eeao}{\end{eqnarray*}}
\newcommand{\be}{\begin{equation}}
\newcommand{\ee}{\end{equation}}
\newcommand{\bea}{\begin{eqnarray}}
\newcommand{\eea}{\end{eqnarray}}
\newcommand{\beq}{\begin{eqnarray}}
\newcommand{\eeq}{\end{eqnarray}}

\def\epem{\ifmmode e^+e^-\else $e^+e^-$\fi}
\def\to{\rightarrow}

\def\mpl{\ifmmode \overline M_{Pl}\else $\bar M_{Pl}$\fi}
\def\beq{\begin{equation}}
\def\be{\begin{equation}}
\def\beqn{\begin{eqnarray}}
\def\ee{\end{equation}}
\def\eeq{\end{equation}}
\def\eeqn{\end{eqnarray}}

\begin{document}

\title{Model-independent constraints on the Abelian $Z'$~couplings within the ATLAS data on the dilepton production processes at $\sqrt{s} =$~13~TeV}

\author{A.~O.~Pevzner}
 \email{apevzner@omp.dp.ua}
\author{V.~V.~Skalozub}
 \email{skalozubv@daad-alumni.de}
\author{A.~V.~Gulov}
 \email{alexey.gulov@gmail.com}
\affiliation{Theoretical Physics Department, Oles Honchar Dnipro
National University, 49010 Dnipro, Ukraine}
\author{A.~A.~Pankov}
\email[]{pankov@ictp.it}
\affiliation{Abdus Salam ICTP Affiliated
Centre at Pavel Sukhoi Gomel State Technical University, Gomel
246746, Belarus}
\affiliation{Institute for Nuclear Problems,
Belarusian State University, Minsk 220030, Belarus}
\affiliation{Joint Institute  for Nuclear Research, Dubna 141980 Russia}

\date{\today}

\begin{abstract}
The study of lepton pair production is a powerful test of  the
Standard Model (SM) and can be used to search for phenomena beyond
the SM. New heavy neutral bosons $Z'$ decaying to charged lepton
pairs $l^+l^-$ ($l=e, \mu$) are predicted by many scenarios of new
physics, including models with extended gauge sector. We estimate the LHC $Z'$ discovery potential with Run
2 data comprised of 36.1 fb$^{-1}$ of $pp$ collisions at
$\sqrt{s}=13$ TeV and recorded by ATLAS detector at the CERN LHC.
The model-independent constraints on the $Z'$ fermion
couplings were obtained for the first time  for a selected set of
$Z'$ signal mass points of $M_{Z'}=2$, 3, and 4 TeV by
using the ATLAS data collected at the LHC. The analysis is based on the derived earlier special relations between the $Z'$ couplings proper to the renormalizable theories. Taking into account the dependence of $Z-Z'$ mixing angle $\theta_0$ on the $Z'$ axial-vector coupling $a$, the limits on $\theta_0$ are established as $\theta_0 < 10^{-4} - 10^{-3}$ in the investigated $Z'$ mass range.

\end{abstract}

 \pacs{12.60.-i}
 \keywords{New gauge $Z'$ bosons,  dilepton resonances, Drell-Yan process,
 ATLAS experiment, Large Hadron Collider (LHC)}

\maketitle

\section{Introduction}

Variety of new physics (NP) scenarios beyond the SM, including
superstring and left-right-symmetric models, predict the existence
of new neutral $Z'$ gauge bosons, which might be light enough to
be accessible at current and/or future
colliders~\cite{Langacker:2008yv,Leike:1998wr,Salvioni:2009mt,Salvioni:2009jp}.
The search for these new neutral $Z^{\prime}$  gauge bosons is an
important aspect of the experimental physics program of current
and future high-energy colliders. Present limits from direct
production at the LHC and virtual (indirect) $Z'$ bounds at LEP,
through interference with the $Z$ boson, imply that new
$Z^{\prime}$ bosons are rather heavy. Depending on the considered
theoretical models, the current limits on $Z^{\prime}$ masses from
their direct search at the LHC at 13 TeV are of the order of
3.7--4.5~TeV \cite{Aaboud:2017buh,CMS:2015nhc}. Another important
dynamical characteristic of the $Z'$ bosons is the $Z-Z'$ mixing
angle which is strongly constrained by the ATLAS and CMS data in
the diboson channel \cite{Osland:2017ema,Pankov:2016hun}. The
diboson decay modes of $Z'$ directly probe the gauge coupling
strength between the new and the SM gauge bosons.

Among the extensions of the SM at the TeV scale, those with an
additional $\tilde{U}(1)$ factor in the gauge group, associated
with a heavy neutral gauge boson $Z'$, have often been considered
in direct and indirect searches for new physics, and in the
studies of possible early and current discoveries at the LHC
\cite{Patrignani:2016xqp}. Many varieties of $Z'$ models have been
considered over the years
\cite{Langacker:2008yv,Leike:1998wr,Salvioni:2009mt,Salvioni:2009jp}.
From now on,  we will concentrate on a class of {\em Abelian}
models, previously discussed in
\cite{Gulov:2000eh,Gulov:2001sg,Gulov:2006gv,Gulov:2011yi,Gulov:2012zz,Gulov:2013dia,Pevzner:2015vwn}.

By \emph{Abelian} $Z'$ models we mean the $\tilde{U}(1)$
extensions of the SM. We assume also that the model contains (i) no exotic vectors, apart from a single $Z^\prime$ associated with an extra $\tilde{U}(1)$
factor in the gauge group, commuting with $SU(3)_C \times SU(2)_L
\times U(1)_Y$, and (ii) no exotic fermions. Abelian
class of $Z'$ models interpolates continuously among several
discrete examples already considered in the literature such as
the $Z'_{\chi}$ model arising from $SO(10)$ unification,
left-right symmetric models, etc.

This class of the Abelian $Z'$ models is motivated
to emerge as a natural benchmark for comparing direct and indirect
signals in different experimental contexts in particular for
organizing experimental searches at the LHC
\cite{Aad:2009wy,cmse,cmsmu}. Nevertheless, the model-dependent $Z'$ analysis may have some difficulties in the nearest future. The identification reach for the majority of the models is about the current estimated lower bound of the $Z'$ mass. This means it will be problematic to distinguish the  basic  $Z'$ model even if the $Z'$ is discovered experimentally.

In such a  situation, a model-independent approach is also very perspective. In contrast to the model-dependent searches for the $Z'$ boson at the LHC
where only one free parameter exists (the mass of the
$Z'$) for a given $Z'$ model, in model-independent approach all fermion
coupling constants are considered as free parameters. Therefore, model-independent
approaches are prospective for estimating not only the mass but also $Z'$ couplings
to the SM particles. As a result, definite classes of the extended models could be restricted.
The obvious shortcoming of model-independent searches is a sufficiently large
amount of free parameters which must be fitted in experiments. However, it can be reduced if
some natural requirements or theoretical arguments are imposed. For example,
the considered Abelian $Z'$ boson assumes that the extended model is a renormalizable one.
This property results in a series of specific relations between couplings called in what follows
the renormalization group relations (RGR), that reduces the number of unknown parameters
essentially. The $Z$--$Z'$ mixing angle becomes also a self-consistent part of the parametric
space. RGR lead to the kinematical structure of the differential cross sections allowing picking out
uniquely the Abelian $Z'$ between other spin-1 neutral particles. In the Abelian class of $Z'$
models, the main $Z'$ properties and its couplings to the SM states can be completely
described in terms of three independent fermion couplings and the $Z'$ mass, $M_{Z'}$.

As we will see below, the parameterization of the Abelian $Z'$ boson is a kind of model-independent
study, which does not specify numeric values of the $Z'$ couplings and automatically takes into account $Z-Z'$ mixing
effects. Such an approach was developed in \cite{Gulov:2000eh,Gulov:2001sg,Gulov:2006gv,Gulov:2011yi,Gulov:2012zz,Gulov:2013dia,Pevzner:2015vwn}.
In these investigations, the mixing angle, the couplings of the $Z'$ to axial and vector lepton and quark currents,
as well as to Higgs field, and the mass $M_{Z'}$ have been estimated at $1\sigma - 2\sigma$ C.L.
mainly at non-resonant energies. In the off-resonanse case the cross section is dominated by the interference of
the SM and the $Z'$ parts in the scattering amplitude. The other terms are next-to-leading in coupling constants and could
be neglected. In this regime, the width of the $Z'$ boson is not very essential.

In all the model-dependent searches fulfilled at the LHC the narrow width approximation has been applied.
It implies that the ratio of the resonance width $\Gamma_{Z'}$  to the $Z'$ mass should be small,
$\Gamma_{Z'}/M_{Z'}\leq 3\%$. This value is typical for the couplings in the usually considered models.
In this case, the interference term in the cross-section is suppressed and the cross section can be approximated as
the product of the $Z'$ production cross section and the $Z'$ decay branchings into the SM particles.

Since the different terms of the cross sections are dominant in
these two ways of data treating, it is of interest to estimate the
parameters of the $Z'$ in the model-independent approach and the
direct search method for the Drell-Yan process within data
accumulated  at $\sqrt{s}= $ 13 TeV \cite{Aaboud:2017buh}. Then it is
interesting to compare them with the ones obtained already in
\cite{Pevzner:2015vwn}, \cite{Gulov:2013dia} at $\sqrt{s} =$~7,~8~ TeV in the indirect search technique. This is the goal of the present paper.

In this work, we derive model-independent bounds on the fermion
$Z'$ coupling constants in dilepton production process
\begin{equation}\label{procll}
p+p \to Z'\to l^+ l^-+X
\end{equation}
($l=e, \mu$) from the available ATLAS  data  at the LHC at
$\sqrt{s}=13$ TeV \cite{Aaboud:2017buh}. We will show that the use
of a model-independent parameterization, such as the one suggested
in present work, is a valuable tool to systematically organize the
$Z'$ searches.

The paper is organized as follows. In
the next section, the information on the RGR necessary for what
follows is given and the
 effective  Lagrangian describing interaction of the $Z'$ with the
  SM particles is adduced. Finally, in the Section 3, the results of the
  $Z'$ couplings estimates are introduced and further are compared with
  some previous known results in the Discussion.

 \section{Effective Lagrangian and RGR relations}
 The effective low energy  Lagrangian
describing the interaction of the $Z$ and $Z'$ mass eigenstates
can be written as (see, e.g. \cite{Gulov:2000eh}):
\begin{multline}\label{ZZplagr}
{\mathcal{L}}_{Z\bar{f}f}=
Z_\mu\bar{f}\gamma^\mu[(v^\mathrm{SM}_{fZ}+\gamma^5
a^\mathrm{SM}_{fZ})\cos\theta_0 \\
 +(v_f+\gamma^5 a_f)\sin\theta_0]f,
\end{multline}
\begin{multline} \label{ZZ'plagr}
{\mathcal{L}}_{Z'\bar{f}f}=
Z'_\mu\bar{f}\gamma^\mu[(v_f+\gamma^5 a_f)\cos\theta_0 \\
- (v^\mathrm{SM}_{fZ}+\gamma^5a^\mathrm{SM}_{fZ})\sin\theta_0]f,
\end{multline}
where $f$ is an arbitrary SM fermion state; $v^\mathrm{SM}_{fZ}$,
$a^\mathrm{SM}_{fZ}$ are the SM axial-vector and vector couplings
of the $Z$-boson, $a_f$ and $v_f$ are the ones for the $Z'$, $\theta_0$ is the
$Z$--$Z'$ mixing angle. The definitions here are such that
\begin{multline} \label{couplings_SM}
v_f^{SM}=(T_{3,f}-2Q_f\hskip 2pt s_W^2)/(2s_Wc_W), \\
a_f^{SM}=-T_{3,f}/(2s_Wc_W).
\end{multline}
In what follows, we will also use the
``normalized'' couplings, \be \label{normcouplings} \bar{a}_f \equiv
\frac{1}{\sqrt{4\pi}} \frac{M_{Z}}{M_{Z'}} \,a_f, \quad \bar{v}_f
\equiv \frac{1}{\sqrt{4\pi}} \frac{M_{Z}}{M_{Z'}}\, v_f. \ee
Within the considered formulation, this angle   is determined by
the coupling $\tilde{Y}_\phi$ of fermions to the scalar field as
follows (see \cite{Gulov:2000eh} and Appendix B of
\cite{Pevzner:2015vwn} for details)
\begin{equation}\label{MixingAngle}
\theta_0 =
\frac{\tilde{g}\sin\theta_W\cos\theta_W}{\sqrt{4\pi\alpha_\mathrm{em}}}
\frac{M^2_Z}{M^2_{Z'}} \tilde{Y}_\phi
+O\left(\frac{M^4_Z}{M^4_{Z'}}\right),
\end{equation}
where $\theta_W$ is the SM Weinberg angle, $\tilde{g}$ is
$\tilde{U}(1)$ gauge coupling constant and $\alpha_\mathrm{em}$ is
the electromagnetic fine structure constant. Although $\theta_0$
is small quantity of order ($m^{2}_{Z}/m^{2}_{Z'}$), it
contributes to the $Z$ and $Z'$ bosons exchange amplitudes and
cannot be neglected.

As  was shown in \cite{Gulov:2000eh,Gulov:2001sg}, if the extended
model is renormalizable and contains the SM as a subgroup, the
relations between the couplings hold:
\be \label{rgrav} v_f - a_f = v_{f^*} - a_{f^*}, \quad a_f =
T_{3f}\tilde{g}\,\tilde{Y}_\phi. \ee Here $f$ and $f^*$ are the
partners of the $SU(2)_L$ fermion doublet ($l^* = \nu_l, \nu^* =
l, q^*_u = q_d$ and $q^*_d = q_u$), $T_{3f}$ is the third
component of the weak isospin.
 These relations  are proper to the models of the Abelian $Z'$.
 They are just  as the correlations for the special  values of the
hypercharges $Y^R_f, Y^L_f, Y_\phi$  of the left-handed,
right-handed fermions, and scalars in the SM.  But now  these
parameters are some arbitrary numbers.

The correlations \Ref{rgrav}  have been already derived in two
different ways. The first one origins from the structure of the
  renormalization group operator and other one is founded on the
 requirement of the SM Yukawa term  invariance with
 respect to the additional $\tilde{U}(1)$ group.

From now on  we will assume that the axial-vector coupling $a_f$
is universal, so that \be \label{a_universality} a \equiv a_e =
-a_{\nu_e} = a_d = -a_u =... \ee Being combined with
Eqs.~(\ref{MixingAngle}) and \Ref{rgrav}, it yields \be
\label{MixingAngle1} \theta_0 = -2a \frac{\sin\theta_W
\cos\theta_W}{\sqrt{4\pi\alpha_{em}}}\frac{M^2_Z}{M^2_{Z'}} +
O\left(\frac{M^4_Z}{M^4_{Z'}}\right). \ee

Eq.~(\ref{rgrav}) plays crucial role in what follows. First, due
to them the number of independent parameters is considerably
reduced. Second (but not less important) is influence on
kinematics of scattering processes. If the signal is observed, due
to these relations of Eq.~(\ref{rgrav}) one can  guarantee  that
exactly the Abelian $Z'$ state is observed.  Finally, it is important to notice that the relations (\ref{rgrav}) hold also in the Two-Higgs-Doublet Model (THDM). All this
makes the direct searching for the $Z'$ combined with \Ref{rgrav}
grounded and perspective.

\section{Numerical analysis  and constraints on $Z'$ couplings}
The differential cross section for $Z^{\prime}$ production in the
process (\ref{procll}) from initial quark-antiquark states can be
written as
\begin{eqnarray}
&& \frac{d\sigma^{Z^\prime}}{dM_{ll}\,dY\,dz}
 = K \frac{2 M_{ll}}{s} \times \\ \nonumber
&&\times\sum_q [f_{q|P_1}(\xi_1)f_{\bar q|P_2}(\xi_2) + \{q \leftrightarrow \bar{q} \}]\, \frac{d\hat \sigma_{q \bar
q}^{Z^\prime}}{dz}. \label{dsigma}
\end{eqnarray}
Here, $s$ denotes the proton-proton center-of-mass energy squared,
$z\equiv\cos\theta$, with $\theta$ the $l^-$--quark angle in the
$l^+l^-$ center-of-mass frame and $Y$ is the dilepton rapidity.
Furthermore, $f_{q|P_1}(\xi_{1},M_{ll})$ and $f_{\bar
q|P_2}(\xi_{2},M_{ll})$ are parton distribution functions for the
protons $P_1$ and $P_2$, respectively, with
$\xi_{1,2}=(M_{ll}/\sqrt s)\exp(\pm Y)$ the parton fractional
momenta. Finally, ${d\hat \sigma_{q \bar q}^{Z^\prime}}/{dz}$ are
the partonic differential cross sections. In~(\ref{dsigma}), the
$K$ factor (=1.3) accounts for higher-order QCD contributions and
we use improved Born approximation for EW part
\cite{Pevzner:2015vwn}.  For numerical computation, we use MSTW
PDF sets \cite{Martin:2009iq}.

The cross section for the narrow $Z'$ state production and
subsequent decay into a $l^+l^-$ pair needed in order to estimate
the expected number of $Z'$ events, $N^{Z^\prime}$, is derived
from (\ref{dsigma}) by integrating the right-hand-side over $z$,
over the rapidity of the $l^\pm$-pair $y$ and invariant mass
$M_{ll}$ around the resonance peak $(M_{Z'}-\Delta M_{ll}/2,$ $
M_{Z'}+\Delta M_{ll}/2)$:
\begin{eqnarray}
&& \sigma^{Z^\prime}{(pp\to l^+l^- + X)}  =\int_{M_{Z'}-\Delta M_{ll}/2}^{M_{Z'}+\Delta M_{ll}/2}d M_{ll} \times \\ \nonumber
&& \times\int_{-Y}^{Y}d Y
\int_{-z_{\text{cut}}}^{z_\text{cut}}d
z\frac{d\sigma^{Z^\prime}}{d M_{ll}\, d Y\, d z}\;, \label{TotCr}
\end{eqnarray}
where the phase space can be found, e.g. in \cite{Osland:2009tn}.
Here, $\Delta{M_{ll}}$ being the mass window. Notice, that in all
the cases studied, the true width of the resonance $\Gamma_{Z'}$
is  smaller than the Gaussian experimental resolution $\Gamma_m$.
For each mass point, having determined the observed resonance
width $\Gamma_m$, a mass window $\Delta{M_{ll}}$ can be defined as
$\pm 3\Gamma_m$ around the $Z'$ mass \cite{Osland:2009tn}.

Using Eq.~(\ref{TotCr}), the number of signal events for a narrow
$Z'$ resonance state can be written as follows
\begin{eqnarray}
&& N^{Z^\prime}= {\cal L}\cdot\varepsilon\cdot
\sigma^{Z^\prime}{(pp\to l^+l^- + X)} \equiv \\ \nonumber
&& \equiv {\cal
L}\cdot\varepsilon\cdot A_{ll}\cdot \sigma(pp\to Z') \times
\text{B}(Z' \to l^+l^-).
 \label{signal}
\end{eqnarray}
Here, ${\cal L}$ denotes the integrated luminosity, and the
overall kinematic and geometric acceptance times trigger,
reconstruction and selection efficiencies,
$A_{ll}\times\varepsilon$, is defined as the number of signal
events passing the full event selection divided by the number of
generated events. Finally, $\sigma(pp\to Z') \times \text{B}(Z'
\to l^+l^-)$ is the (theoretical) total production cross section
times branching ratio extrapolated to the total phase space.
Finally,  $\text{B}(Z' \to l^+l^-)=\Gamma(Z' \to
l^+l^-)/\Gamma_{Z'}$ where $\Gamma(Z' \to l^+l^-)$ and
$\Gamma_{Z'}$ being the partial lepton and total widths of the
$Z'$ boson.

In the calculation of the total width $\Gamma_{Z'}$ we included
the following channels: $Z'\to f\bar f$, $W^+W^-$, and $ZH$
\cite{Osland:2017ema}, where $H$ is the SM Higgs boson and $f$ are
the SM fermions ($f=l,\nu,q$). The total width $\Gamma_{Z'}$ of
the $Z'$ boson can be written as  follows:
\begin{equation}\label{gamma2}
\Gamma_{Z'} = \sum_f \Gamma_{Z'}^{ff} + \Gamma_{Z'}^{WW} +
\Gamma_{Z'}^{ZH}.
\end{equation}
The presence of the two last decay channels,  which are often
neglected, is due to $Z$-$Z'$ mixing. However for large $Z'$
masses there is an enhancement that cancels the suppression due to
tiny $Z-Z'$ mixing \cite{Salvioni:2009mt}.  The ratio
$\Gamma_{Z^\prime} / M_{Z'}$ is pretty constant over the whole
range of masses of interest, and is around 2\% for representative
$Z'$ models originated from $E_6$  GUT and LR scenarios. Notice
that for all $M_{Z'}$ values of interest for LHC the width of the
$Z'$ boson should be considerably smaller than the mass window
$\Delta M_{ll}$ in order to  meet the narrow width approximation
(NWA) condition.

In  early study \cite{Gulov:2013dia}, to gain some approximate
understanding of the acceptances for signal and background at
different values of the invariant mass $M_{ll}$ of the $l^+l^-$
pair and $Z'$ mass, we performed a simple study as follows. To
estimate the 2$\sigma$ constraints on the $Z'$ parameters at the
LHC, we compared the events due to a $Z'$ signal to the events
from the SM background in a 3$\%$ interval around the relevant
values of the dilepton $l^+l^-$ invariant mass. This should be
compatible with the expected energy resolution and with the fact
that $\Gamma_{Z'} / M_{Z'} < 3.0 \%$. We then required the signal
events to be at least a $2\sigma$ fluctuation over the expected
background, and in any case more than 3. This rough statistical
analysis, as a preliminary stage, was enough to get an approximate
answer to the questions we wanted to address.

Here, we are making a more careful analysis, employing the most
recent measurements of dilepton processes provided by the
experimental collaboration ATLAS,  which have control on all the
information needed to perform it in a more accurate way. In
particular, for Abelian $Z'$ we compute the LHC $Z'$ production
cross-section multiplied by the branching ratio into two leptons
$l^+l^-$, $\sigma(pp\to Z') \cdot {\rm B}(Z'\to l^+ l^-)$, as a
function of three free parameters (${a},{v}_e,{v}_u$) at given
$Z'$ mass $M_{Z'}$ and compare it with the limits of $\sigma_{\rm
95\% CL}\cdot B$  obtained from ATLAS data.

\begin{figure}[h!]
    \caption{Observed and expected $95\%$ C.L. upper limits $\sigma_{\rm 95\% CL}\cdot B$ on the
production cross section times the branching fraction for $Z'\to
l^+l^-$ as a function of $Z'$ mass, $M_{Z'}$  obtained from ATLAS
data for $36.1~\text{fb}^{-1}$ \cite{Aaboud:2017buh}. The inner
(green) and outer (yellow) bands around the expected limits
represent $\pm 1\sigma$ and $\pm 2\sigma$ uncertainties,
respectively.  Theoretical production cross sections $\sigma\cdot
 B(Z'\to l^+l^-)$ for Abelian $Z'$ and some representative models
($Z'_\chi$, $Z'_\psi$ and $Z'_{\rm SSM}$) are calculated with
FeynArts and FormCalc \cite{Kublbeck:1992mt} with a $K$-factor of
1.3. The allowed (excluded) signature space for Abelian $Z'$ lies
below (above) the upper limits $\sigma_{\rm 95\% CL}\cdot B$. The
former ones are indicated as vertical lines for three
representative values $M_{Z'}=2$, 3, and 4 TeV.}
    \centering
    {\includegraphics[scale=1]{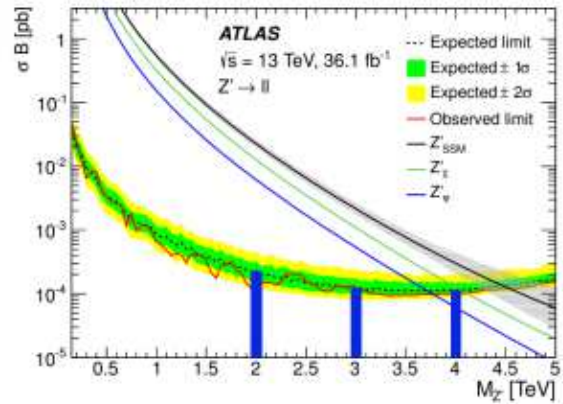}}
    \label{fig1}
\end{figure}

Our strategy in the present analysis is to use the SM backgrounds
that have been carefully evaluated by the experimental
collaboration and we simulate only the $Z^\prime$ signal.
Fig.~\ref{fig1} shows the observed and expected $95\%$ C.L. upper
limits on the production cross section times the branching
fraction for $Z'\to l^+l^-$ as a function of $Z'$ mass, $M_{Z'}$.
The data analyzed comprises $pp$ collisions at $\sqrt{s}=13$ TeV,
recorded by the ATLAS (36.1 fb$^{-1}$) detector at the LHC
\cite{Aaboud:2017buh}. The inner (green) and outer (yellow) bands
around the expected limits represent $\pm 1\sigma$ and $\pm
2\sigma$ uncertainties, respectively. Also shown are theoretical
production cross sections $\sigma(pp\to Z')\cdot \rm{B}(Z'\to
l^+l^-)$ for Abelian $Z'$ and some representative models
($Z'_\chi$, $Z'_\psi$ and $Z'_{\rm SSM}$) are calculated with
FeynArts and FormCalc \cite{Kublbeck:1992mt} for  $K$-factor of
1.3. Notice, that the allowed (excluded) signature space for
Abelian $Z'$ lies below (above) the upper limits $\sigma_{\rm 95\%
CL}\cdot B$. The former ones are presented as vertical lines for
three representative values $M_{Z'}=$2 TeV, 3 TeV and 4 TeV.

\begin{figure}[h!]
\includegraphics[scale=0.7]{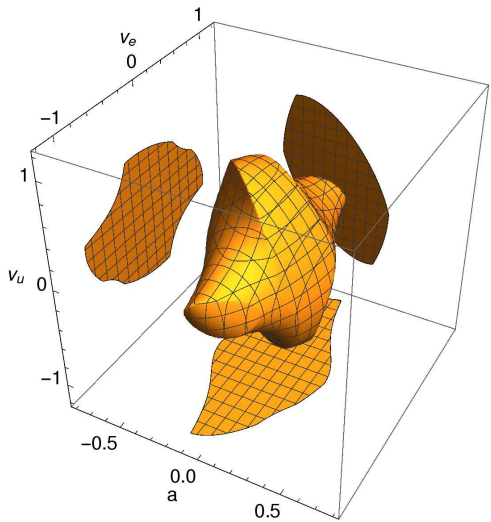}\vspace{0.3cm}
\caption{ 95\% C.L. allowed three-dimensional surface for Abelian
$Z'$ with $M_{Z'}=$3 TeV obtained from the comparison of the
theoretical cross section $\sigma(pp\to Z') \cdot {\rm B}(Z'\to
l^+ l^-)$ vs $\sigma_{\rm 95\% CL}\cdot B$ from dilepton
production ATLAS data. Also, the condition $\Gamma_{Z'} / M_{Z'} <
1.0 \%$ was taken into account.
Two-dimensional projections of allowed region are also shown.
  } \label{fig2}
\end{figure}

This procedure gives the upper constraints on the $Z'$ couplings. As mentioned above, due to the relations (\ref{rgrav}) the cross section of the process $pp \to Z' \to l^+ l^- + X$ can be expressed through $a$, $v_e$, $v_u$ couplings only. Hence any constraints on the $Z'$ production cross section at a given $M_{Z'}$ yield the corresponding constraints on these couplings. In Fig.~\ref{fig2} we displays in 3-dim parameter
space (${a},{v}_e,{v}_u$) upper model-independent bounds at 95\%
C.L. on $Z'$ parameters  obtained from equation $\sigma(pp\to Z')
\cdot {\rm B}(Z'\to l^+ l^-)=\sigma_{\rm 95\% CL}\cdot B$ at
$M_{Z'}=$3 TeV. In Fig.~\ref{fig2}  we also show the planar
regions that are obtained by projecting onto the 2-dim planes
 the 95\% C.L. allowed three-dimensional surface.

\begin{figure}[h!]
    \caption{The 95\% C.L. allowed areas in the
$(a, v_e)$ parameter space, derived from the cross-section limits
for three sample $Z'$ masses 2 TeV, 3 TeV and 4 TeV. The
hyperbolic regions correspond to the condition of $\Gamma_{Z'} /
M_{Z'} < 0.01$ and $<0.03$. Parameters for the representative
models are also shown.
    }
    \centering
    {\includegraphics[scale=1]{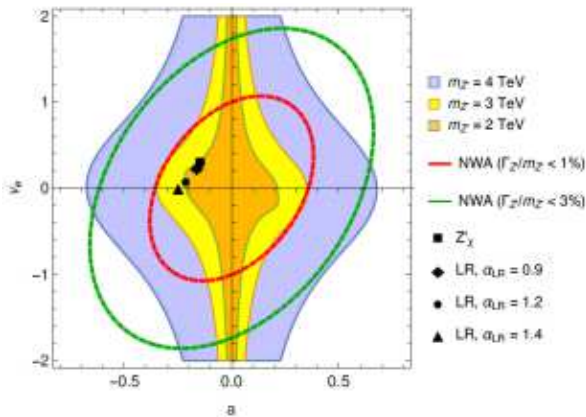}}
    \label{fig3}
\end{figure}

\begin{figure}[h!]
    \caption{Same as in Fig.~\ref{fig3} but for $(a, v_u)$ parameter space.}
    \centering
    {\includegraphics[scale=1]{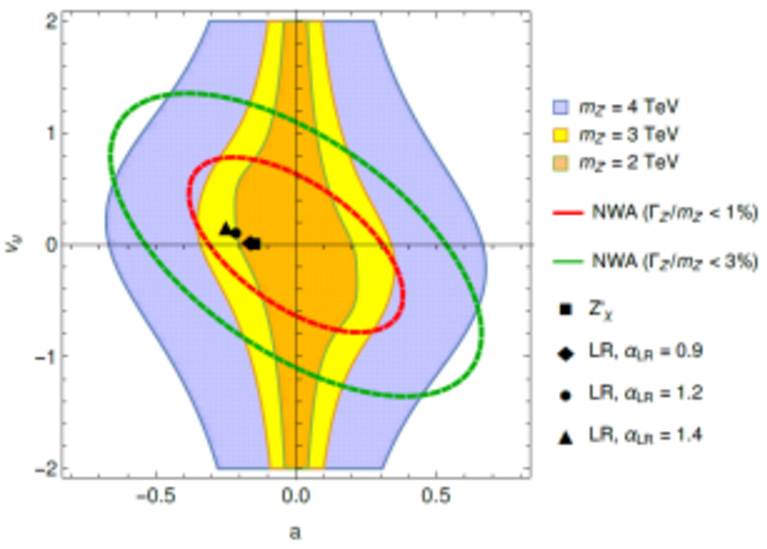}}
    \label{fig4}
\end{figure}

\begin{figure}[h!]
    \caption{Same as in Fig.~\ref{fig3} but for $(v_e, v_u)$ parameter space.}
    \centering
    {\includegraphics[scale=1]{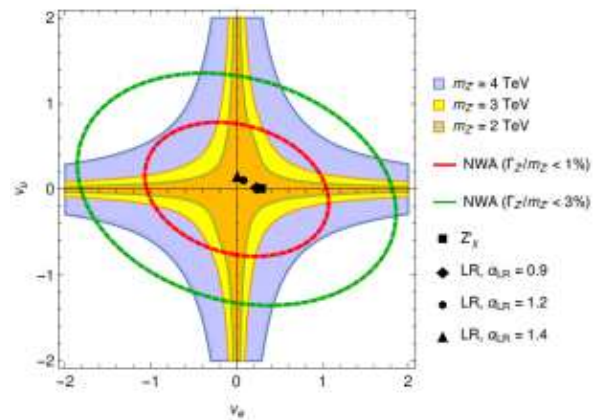}}
    \label{fig5}
\end{figure}

The Fig.~\ref{fig3} shows the exclusion contours at 95\% C.L. in the
$(a, v_e)$ parameter space, derived from the cross-section limits
for three sample $Z'$ masses 2 TeV, 3 TeV and 4 TeV. The region
inside each ellipse indicates the part of the parameter space in
which the ratio of the resonance total width $\Gamma_{Z'}$ to its
mass $M_{Z'}$ is below 1\% (or 3\%), which is comparable to the
experimental mass resolution.  Parameters for the benchmark models
(${Z'}_{\chi}$ and ${Z'}_{\rm LR}$) are also shown by the the
model marks (circle, rhombus, square, and triangle). The solid lines
bounding the allowed areas indicated by color represent the
boundaries of the regions excluded by this search for different
$Z'$  masses (the region outside these lines is excluded). Points
inside the ellipse, but where the absolute values of the $a$ and
$v_e$ parameters are larger than at the exclusion contour, are
considered to be excluded at a C.L. greater than 95\%.
Fig.~\ref{fig4} and Fig.~\ref{fig5} show similar exclusion
contours at 95\% C.L. in the $(a, v_u)$ and  $(v_e, v_u)$
parameter space, respectively.

\section{Discussion and concluding remarks}
In one of our previous work \cite{Pevzner:2015vwn}, we carried out
the model-independent analysis of $Z'$ effects in its indirect
search at the LHC. It means that we supposed that the $Z'$ has a
mass beyond the LHC reach.  In such an approach, the $Z-Z'$
interference terms (which are proportional to the quadratic $Z'$
couplings) dominate in the $Z'$ production cross section
$\sigma_{Z'}$. Besides, in this case the $Z'$ decay width
$\Gamma_{Z'}$ is almost unimportant in the analysis of inderect
(interference) $Z'$ effects. The indirect search allowed us to
establish the maximum likelihood estimations on the $Z'$ couplings
\cite{Pevzner:2015vwn} as $\bar{a}^2 \sim 10^{-5}$,
$|\bar{a}\bar{v}_e| \sim 10^{-6}$, $|\bar{a}\bar{v}_u| \sim
10^{-3}$ in the $Z'$ mass range 1200~GeV~$< M_{Z'} <$~4500~GeV at
$\sqrt{s}=7$ and 8 TeV.
\begin{table}[ht]
    \caption{Model-independent  upper limits at 95\% C.L. on fermion couplings based on
    ATLAS dilepton production
 data in  direct $Z'$ search at 13 TeV} \label{tbl:Constraints}
    \centering
    \begin{tabular}{|c|c|c|c|c|}
        \hline
        & $M_{Z'} = 2$~ TeV & $M_{Z'} = 3$~ TeV & $M_{Z'} = 4$~ TeV \\ \hline
        $|a|$ & 0.20 & 0.35 & 0.35 \\ \hline
        $|\bar{a}|$ & 0.003 & 0.002 & 0.002 \\ \hline \hline
        $|v_e|$ & 1 & 1.1 & 1.1 \\ \hline
        $|\bar{v}_e|$ & 0.01 & 0.007 & 0.005 \\ \hline\hline
        $|v_u|$ & 0.8 & 0.9 & 0.9 \\ \hline
        $|\bar{v}_u|$ & 0.01 & 0.007 & 0.005 \\ \hline

    \end{tabular}
\end{table}

Nevertheless, we cannot put any definite assumptions on $M_{Z'}$
until the $Z'$ is detected explicitly. Thus the direct $Z'$ search
must be also performed as well as the indirect one. Within the
direct search concept we believe each time that the energy of the
experiment is close to the $Z'$ mass. Hence the $Z-Z'$
interference terms turn out to be suppressed in the cross section
while the pure $Z'$ production part (proportional to the quartic
$Z'$ couplings) dominates. What about  $\Gamma_{Z'}$, its value
becomes crucial for the direct search. It is often believed that
we deal with a narrow peak, $\Gamma_{Z'} / M_{Z'} \leq 0.01-0.03$.

The present paper is based on the ATLAS data on the Drell-Yan
production at $\sqrt{s}=13$ TeV. The upper constraints on the $Z'$
couplings were obtained as a result of applying two conditions.
The first one is a comparison of resonant production cross section
with $95\%$ C.L. upper limits $\sigma_{\rm 95\% CL}\cdot B$. The
second condition is associated with NWA.  We performed the
analysis of direct $Z'$ search allowing to vary $M_{Z'}$ within
the interval 1.25~TeV~$ < M_{Z'} < $~4.5~TeV and obtained that
$(\bar{a}^2)_{\mathrm{max}} \sim 10^{-6}$,
$|\bar{a}\bar{v}_e|_{\mathrm{max}} \sim 10^{-6} - 10^{-5}$,
$|\bar{a}\bar{v}_u|_{\mathrm{max}} \sim 10^{-6}$.

Let us compare these results with some ones known from the
literature (\cite{Gulov:2013dia}, \cite{Osland:2017ema}, \cite{Pankov:2016hun}). At first, they almost
coincide with the $Z'$ constraints derived from the LHC data
analysis at 8 TeV \cite{Gulov:2013dia}. Also, using (\ref{MixingAngle1}) and $|\bar{a}|_{\mathrm{max}}$
from the Table \ref{tbl:Constraints} it is possible to calculate
the upper limit for the $Z-Z'$ mixing angle. It is estimated as
$|\theta_0| < 10^{-4} - 10^{-3}$ for the considered mass range
2~TeV~$ < M_{Z'} < $~4~TeV. This evaluations are consistent
with the results obtained from the global LEP data analysis
\cite{Erler:2009jh} and with direct $Z'$ search in the diboson
channel at the LHC at 13 TeV \cite{Osland:2017ema,Pankov:2016hun}.

In conclusion, we studied  Abelian $Z'$ bosons, whose
phenomenology is controlled by only the $Z^\prime$ mass and three
fermion coupling constants.  We estimated the LHC discovery
potential with Run 2 data comprised of 36.1 fb$^{-1}$ of $pp$
collisions at $\sqrt{s}=13$ TeV and recorded by ATLAS detector at
the CERN LHC. In particular, the model independent limits on the
$Z'$ fermion couplings were obtained for the first time  for
representative $Z'$ signal mass points of $M_{Z'}=2$, 3, and
4~TeV by using the ATLAS data collected at the LHC.

\section*{Acknowledgments}
This research has been partially supported by the Abdus Salam ICTP
(TRIL Programme) and the Belarusian Republican Foundation for
Fundamental Research. One of the authors [A.O.P.] acknowledges the
receipt of the grant from the Abdus Salam International Centre for
Theoretical Physics, Trieste, Italy.

\end{document}